\documentclass[12pt]{article}\usepackage{EntFor+CondInfo}

\begin{document}

\title{Entanglement of Formation\\
 and\\
Conditional Information Transmission}

\author{Robert R. Tucci\\
        P.O. Box 226\\ 
        Bedford,  MA   01730\\
        tucci@ar-tiste.com}

\date{ \today} 

\maketitle

\vskip2cm
\section*{Abstract}
We show that the separability of states in quantum mechanics has a close 
counterpart in classical physics, and that conditional mutual information 
(a.k.a. conditional information transmission) 
is a very useful quantity
in the study of both quantum and classical separabilities.  
We also show how to define
entanglement of formation in terms of conditional mutual information. 
This paper lays the theoretical foundations for a 
sequel paper
which  will present a computer program 
that  can calculate
a decomposition of any separable quantum or classical state.

\newpage
\section{Introduction}

Recently, a few authors\cite{Tucci-tang99}\cite{Gisin}\cite{Tucci-tang00a} have 
noticed a deep connection
between conditional mutual information and quantum separability.
In a parallel development, some researchers\cite{Holland}\cite{Heifei}
have recently 
proven theorems giving necessary and sufficient conditions for 
quantum separability using ideas that hark back to a paper 
by Hughston-Jozsa-Wootters\cite{HJW}. An important goal of this paper
is to tie together these two apparently disconnected lines of thought.

In this paper, we explore the classical roots
of quantum entanglement.
We show that the separability of states in quantum mechanics has a close 
counterpart in classical physics, and that conditional mutual information 
is a very useful quantity
in the study of both quantum and classical separabilities. 

In this paper, we also show how to define entanglement of formation in terms of 
conditional mutual information.

In a sequel paper that will soon follow, we will present a computer program 
based on the theory of this paper. Our software
uses a relaxation algorithm to calculate
a decomposition of any separable quantum or classical state.
The authors of Ref.\cite{Holland} have
written some excellent software 
that can calculate similar things using an algorithm different from ours.

\section{Notation}

In this section, we will introduce certain notation which is 
used throughout the paper. 

For any finite set $S$, let $|S|$ denote the number of elements
in $S$. 
The 
Kronecker delta function $\delta(x,y)$ equals one if $x=y$ and zero otherwise.
We will often abbreviate $\delta(x,y)$ by $\delta^x_y$. 
For any Hilbert space $\hil$,
$dim(\hil)$ will stand for the dimension of $\hil$.
If $\ket{\psi}\in \hil$, then we will often represent 
the projection operator $\ket{\psi}\bra{\psi}$ by $\pi(\psi)$. 

We will underline random variables. For example,
we might write $P(\rvx = x)$ for the probability that
the random variable $\rvx$ assumes value $x$.
$P(\rvx=x)$ will often be abbreviated by $P(x)$ when no 
confusion will arise. $S_\rvx$ will denote the
set of values which the random variable $\rvx$
may assume, and $N_\rvx$ will denote the number of
elements in $S_\rvx$. With each random variable $\rvx$,
we will associate an orthonormal basis $\{ \ket{x} | x\in S_\rvx \}$
which we will call the {\it $\rvx$ basis}. We will 
represent by $\hil_\rvx$ the Hilbert space spanned by the $\rvx$ basis.
Thus, 
$dim \hilx = N_\rvx$.

For any two random variables $\rvx$ and $\rvy$,
$S_{\rvx, \rvy}$ will represent the direct product set 
$S_\rvx \times S_\rvy = \{(x,y) | x\in S_\rvx, y\in S_\rvy\}$. Furthermore, 
$\hil_{\rvx, \rvy}$ will represent $\hil_\rvx \otimes \hil_\rvy$, the tensor product of 
Hilbert spaces
$\hil_\rvx$ and $\hil_\rvy$.
If $\ket{x}$ for all $x$ is the $\rvx$ basis and 
$\ket{y}$ for all $y$ is the $\rvy$ basis, then 
$\hilxy$ is the vector space 
spanned by 
$\{\ket{x,y} | x\in S_\rvx, y\in S_\rvy\}$, where $\ket{x,y} = \ket{x}\ket{y}$.

For any $|\psi_\rvx\rangle \in \hil_\rvx$,
we will use $\psi_x$ to represent  $\av{x|\psi_\rvx}$.
For any $|\psi_{\rvx \rvy}\rangle \in \hilxy$,
we will use $\psi_{xy}$ to represent  $\av{x,y|\psi_{\rvx \rvy}}$.

$\pd(S_\rvx)$ will denote the set of all probability distributions
$P(\cdot)$ for the random variable $\rvx$; i.e., 
all functions $P:S_\rvx\rarrow[0,1]$ such that $\sum_x P(x) = 1$. 
$\dm(\hilx)$ will denote the set of all density matrices acting on the  Hilbert space $\hilx$; i.e., 
the set of all $N_\rvx$ dimensional Hermitian matrices with unit trace and non-negative eigenvalues.

Whenever we use the word ``ditto", as in ``X (ditto, Y)", we mean
that the statement is true if X is replaced by Y. For example,
if we say ``A (ditto, X) is smaller than B (ditto, Y)", we mean 
``A is smaller than B" and ``X is smaller than Y".

This paper will also utilize certain notation
associated with classical and quantum entropy.
See Ref.\cite{Tucci-review} for definitions and examples of the use of such notation.

\section{Classical Separability}

In this section, we will discuss classical separability. In the next section, we will discuss
quantum separability, stressing the similarities with the classical case.
 
 We will say $P(x,y)\in \pd(S_{\rvx\rvy})$ is {\it $N$-separable} iff there exists
 some random variable $\rvalp$ with $N_\rvalp=N$ and there exist probability distributions
 $\tilde{P}(x|\alpha)\in \pd(S_\rvx)$,
 $\tilde{P}(y|\alpha)\in \pd(S_\rvy)$,
 $\tilde{P}(\alpha)\in \pd(S_\rvalp)$
 such that $P(x,y)$ can be ``decomposed" thus:
 
 \beq
 P(x, y) = \sum_\alpha \tilde{P}(x|\alpha) \tilde{P}(y|\alpha) \tilde{P}(\alpha)
 \;.
 \label{eq:clas-separa}\eeq
 We will also say that $P(x,y)\in \pd(S_{\rvx\rvy})$ is {\it separable} 
 iff it is $N$-separable for some $N$.
 
 \begin{theo} \label{th:clas-separ}
 $P(x,y)\in \pd(S_{\rvx\rvy})$ is $N$-separable (ditto, separable) if and only if 
 there exists $\tilde{P}(x, y, \alpha)\in \pd(S_{\rvx \rvy \rvalp})$
 with $N_\rvalp = N$ (ditto, with $N_\rvalp$ arbitrary)
 such that 
 \beq
 P(x, y) = \sum_\alpha \tilde{P}(x, y, \alpha)
 \;
 \label{eq:clas-bound-cond}\eeq
 and
 \beq
 \tilde{P}(x, y, \alpha)\tilde{P}(\alpha) = \tilde{P}(x,\alpha)\tilde{P}(y,\alpha)
 \;
\label{eq:clas-cond-indep}\eeq
for all $(x, y,\alpha)\in S_{\rvx\rvy\rvalp}$.
 (The last condition is just another way of expressing conditional independence:
 \beq
 \tilde{P}(x, y| \alpha) = \tilde{P}(x|\alpha)\tilde{P}(y|\alpha)
 \;.
 \eeq
 )
 \end{theo}
 proof:

 ($\Rightarrow$) Since $P(x,y)$ is separable, there exist probability distributions
$\tilde{P}(x|\alpha)$, $\tilde{P}(y|\alpha)$, $\tilde{P}(\alpha)$. Define 
$\tilde{P}(x, y, \alpha)\in \pd(S_{\rvx \rvy \rvalp})$ by
$\tilde{P}(x, y,\alpha) = \tilde{P}(x|\alpha)\tilde{P}(y|\alpha)\tilde{P}(\alpha)$
$\tilde{P}(x, y,\alpha)$ clearly satisfies all the conditions 
imposed upon it by the right hand side of the theorem.

($\Leftarrow$) The right hand side of the theorem provides us with 
$\tilde{P}(x, y, \alpha)\in \pd(S_{\rvx \rvy \rvalp})$. We can use it to construct 
conditional probabilities  $\tilde{P}(x|\alpha)$,
 $\tilde{P}(y|\alpha)$ and
 $\tilde{P}(\alpha)$ which satisfy Eq.(\ref{eq:clas-separa}). QED
 
 Suppose $f(x,y)$ is a real valued function  of two arguments $x,y$ (i.e., $f:S_\rvx\times S_\rvy\rarrow R$).
  Let $*$ stand for either addition or multiplication.
  If 
 there exist real valued
 functions $f_1(x)$ and $f_2(y)$ such that
 $f(x,y) = f_1(x)*f_2(y)$ for all $x,y$, then we will say  
  that $f$ is
 an {\it x, y *corrugated surface}. 
 
 Suppose that $f(x,y) = f_1(x)*f_2(y)$ is a *corrugated surface such that 
the functions $f_1, f_2$ are
 differentiable.
 Suppose
 the z axis points upward, 
 the x axis eastward, and 
 the y axis northward. If we plot $f(x,y)$ along the z direction, then 
 the mountain tops and valley bottoms 
of the $f$  surface are all oriented along 
either the east-west or
the north-south directions.  Indeed, if at  $x= x_0$, $\partial_xf_1(x_0)=0$, then 
$\partial_xf(x_0,y)=0$ for all $y$; and likewise if $\partial_yf_2(y_0)=0$, then 
$\partial_yf(x,y_0)=0$ for all $x$. This is true regardless of whether * stands for multiplication or addition.

\begin{figure}[h]
	\begin{center}
	\epsfig{file=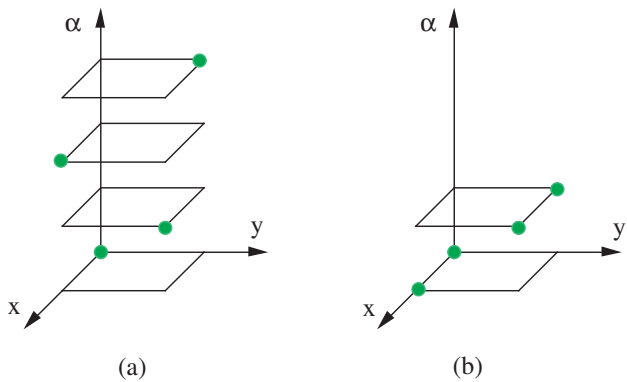, height=2.0in}
	\caption{Two $(x,y,\alpha)$ lattices. Filled circles represent lattice points which have non-zero probability.}
	\label{fig:clas-decompo}
	\end{center}
\end{figure}

Now consider any $\tilde{P}(x, y, \alpha)\in \pd(S_{\rvx \rvy \rvalp})$.
If $\tilde{P}(x, y, \alpha)$ satisfies Eq.(\ref{eq:clas-cond-indep}), 
then it is an $x,y$ product-corrugated surface at fixed $\alpha$.
One can convey this concept graphically by 
drawing a 3-dimensional orthogonal lattice with main axes $x, y, \alpha$,
and writing at each lattice point $(x,y,\alpha)\in S_{\rvx\rvy\rvalp}$ 
the value $\tilde{P}(x, y, \alpha)$.
Each $\alpha\in S_\rvalp$ determines a different horizontal plane.
The values of $\tilde{P}(x, y, \alpha)$ at each horizontal plane are product-corrugated.
This geometrical insight immediately suggest that all
$P(x,y)\in \pd(S_{\rvx\rvy})$  are separable. Two possible decompositions of 
$P(x,y)$ are as follows:

($a$) Suppose that $N_\rvalp= N_\rvx N_\rvy$ and each $\alpha$ plane has a single lattice point
$(x_\alpha, y_\alpha)$ with non-zero probability. Furthermore,
suppose that the point with non-zero probability is different for each $\alpha$ plane (i.e.,
$(x_{\alpha_1}, y_{\alpha_1}) \neq (x_{\alpha_2}, y_{\alpha_2})$ iff $\alpha_1 \neq \alpha_2$.)
See Fig.(\ref{fig:clas-decompo}a) for an example with $N_\rvx = N_\rvy = 2$.
Let $\alpha(x,y)$ be a 1-1 onto function which maps $S_{\rvx \rvy} \rarrow S_{\rvalp}$
and $(x_\alpha, y_\alpha) \rarrow \alpha$. Define $\tilde{P}(\cdot)$ by

\beq
\tilde{P}(x, y, \alpha) = P(x,y) \delta(\alpha, \alpha(x,y)) 
= P(x_\alpha, y_\alpha)\delta(x, x_\alpha) \delta(y, y_\alpha)
\;.
\eeq
It is easy to check that $\tilde{P}(\cdot)$ is an element of $\pd(S_{\rvx \rvy \rvalp})$
that satisfies Eqs. (\ref{eq:clas-bound-cond}) and (\ref{eq:clas-cond-indep}).

($b$) Suppose that $N_\rvalp= N_\rvy$ and at each $\alpha$ plane
all lattice points have zero probability except for possibly
those in a line of lattice points
$L_\alpha= \{ (x, y_\alpha) | x\in S_\rvx \}$. Furthermore, suppose
$L_{\alpha_1} \neq L_{\alpha_2}$ iff $\alpha_1 \neq \alpha_2$.
See Fig.(\ref{fig:clas-decompo}b) for an example with $N_\rvx=N_\rvy = 2$.
Let $\alpha(y)$ be a 1-1 onto function which maps $S_{\rvy} \rarrow S_{\rvalp}$
and $y_\alpha \rarrow \alpha$. Define $\tilde{P}(\cdot)$ by

\beq
\tilde{P}(x, y, \alpha) = P(x,y) \delta(\alpha, \alpha(y)) 
= P(x, y_\alpha)\delta(y, y_\alpha)
\;.
\eeq
It is easy to check that this $\tilde{P}(\cdot)$ is an element of $\pd(S_{\rvx \rvy \rvalp})$
that satisfies Eqs. (\ref{eq:clas-bound-cond}) and (\ref{eq:clas-cond-indep}).

Note that even though every $P(x,y)\in \pd(S_{\rvx\rvy})$ is separable, it may not be 
$N$-separable. Example ($b$) above implies that $P(x,y)$ is 
$N_\rvalp$ separable when $N_\rvalp\geq min(N_\rvx, N_\rvy)$, but what if $N_\rvalp$ is smaller than this?
(for example, if $N_\rvalp =2$ but $N_\rvx, N_\rvy >> 2$). For small enough $N_\rvalp$, it may be impossible 
to construct a $\tilde{P}(x, y, \alpha)$ that satisfies all the constraints 
given by  Eqs. (\ref{eq:clas-bound-cond}) and (\ref{eq:clas-cond-indep}).

For $P(x,y)\in \pd(S_{\rvx\rvy})$ and any integer $N\geq 1$, define

\beq
E^{(N)}_F(P) = \frac{1}{2} \min_{\tilde{P}} H(\rvx: \rvy | \rvalp)
\;,
\eeq
where the minimum is taken over the set of all $\tilde{P}(x,y,\alpha)\in \pd (S_{\rvx\rvy\rvalp})$
such that $N_\rvalp = N$ and $P(x,y) = \sum_\alpha \tilde{P}(x,y,\alpha)$. The conditional 
mutual entropy $H(\rvx: \rvy | \rvalp)$ is calculated for the probability distribution $\tilde{P}$.

\begin {theo}
$P(x,y)\in \pd(S_{\rvx\rvy})$ is $N$-separable if and only if $E^{(N)}_F(P) = 0$.
\end{theo}
proof:

($\Rightarrow$) Clear.

($\Leftarrow$) There exists a $\tilde{P}(x,y,\alpha)\in \pd(S_{\rvx\rvy\rvalp})$ such 
that $N_\rvalp =  N$, $P(x,y) = \sum_\alpha \tilde{P}(x,y,\alpha)$, and 
$H(\rvx: \rvy | \rvalp)=0$. Because this conditional mutual entropy vanishes, 
 $\tilde{P}(x,y|\alpha) = \tilde{P}(x|\alpha)\tilde{P}(y|\alpha)$. Hence,
$P(x,y)$ is $N$-separable. QED

\section{Quantum Separability}

We say $\rho \in \dm(\hilxy)$ is {\it $N$-separable} (ditto, {\it separable}) 
iff there exist a random variable $\rvmu$ with 
$N_\rvmu=N$ (ditto, with arbitrary $N_\rvmu$) and there exist
$P(\mu) \in \pd(S_\rvmu)$, 
$\rho_\rvx^\mu \in \dm(\hilx)$ and 
$\rho_\rvy^\mu \in \dm(\hily)$ 
such that $\rho$ can be ``decomposed" thus:

\beq
\rho = \sum_\mu P(\mu) \rho_\rvx^\mu \rho_\rvy^\mu
\;.
\eeq

An  equivalent definition is: $\rho \in \dm(\hilxy)$ is {\it $N$-separable} (ditto, {\it separable}) 
iff there exist a random variable $\rvalp$ with 
$N_\rvalp=N$ (ditto, with arbitrary $N_\rvalp$) and there exist 
$w_\alpha \in pd(S_\rvalp)$,
$|\psi^\alpha_\rvx\rangle\in \hilx$ and
$|\psi^\alpha_\rvy\rangle\in \hily$
such that $\rho$ can be ``decomposed" thus:

\beq
\rho = \sum_\alpha w_\alpha \pi(\psi^\alpha_\rvx) \pi(\psi^\alpha_\rvy)
\;.
\label{eq:quan-separ-def}\eeq

The second definition clearly implies the first. To see that the first definition
implies the second: for each $\mu$, express $\rho_\rvx^\mu $ and
$\rho_\rvx^\mu $ in terms of their eigenstates:

\beq
\rho_\rvx^\mu = \sum_a P(a|\mu) \ket{\phi^{\mu a}_\rvx}\bra{\phi^{\mu a}_\rvx}
\;,
\eeq

\beq
\rho_\rvy^\mu = \sum_b P(b|\mu) \ket{\phi^{\mu b}_\rvy}\bra{\phi^{\mu b}_\rvy}
\;.
\eeq
We identify the index $\alpha$ with the 3-tuple $(\mu, a, b)$ so $S_\rvalp = S_{\rvmu \rva\rvb}$.
We define for all $\alpha\in S_\rvalp$:

\beq
w_\alpha = P(a|\mu) P(b|\mu) P(\mu)
\;,
\eeq

\beq
\ket{\psi^\alpha_\rvx} = \ket{\phi^{\mu a}_\rvx}
\;
\eeq
(note that the right hand side is the same for all $b$), and

\beq
\ket{\psi^\alpha_\rvy} = \ket{\phi^{\mu b}_\rvy}
\;
\eeq
(note that the right hand side is the same for all $a$).
With these definitions, Eq.(\ref{eq:quan-separ-def}) follows. 

In the previous section about classical separability, we 
encountered several 
theorems of the form:
``$P$ is separable iff condition X".
These theorems had matching theorems of the form
``$P$ is $N$-separable iff $N=N_\rvalp$ and condition X".
In what follows, we will often
encounter theorems of the form:
``$\rho$ is separable iff condition X".
As in the classical case, these theorems 
about separability have obvious matching theorems about 
$N$-separability
(``$\rho$ is $N$-separable iff $N=N_\rvalp$ and condition X"), but 
for simplicity, we will not mention them henceforth.

Consider some Hilbert space $\hil$ and some $\rho\in \dm (\hil)$.
$\rho$ can be expressed as

\beq
\rho = \sum_j \lambda_j | \phi^j \rangle\langle \phi^j |
\;,
\label{eq:evec-expan}\eeq
where $(\lambda_j , | \phi^j\rangle)$ for all $j$ are
the eigenvalues and eigenvectors  of $\rho$. 
In Ref.\cite{HJW}, Hughston, Jozsa and Wootters (HJW) proved the following theorem.

\begin{theo}[HJW]\label{th:HJW}
$\rho\in \dm(\hil)$ can be expressed as

\beq
\rho = \sum_\alpha w_\alpha | \psi^\alpha \rangle\langle \psi^\alpha |
\;,
\label{eq:alpha-expan}\eeq
where $w_\alpha \in \pd(S_\rvalp)$, and 
$\ket{\psi^\alpha}\in \hil$ for all $\alpha$
if and only if there exists a transformation $T^\alpha_j$ 
($\alpha \in S_\rvalp$, $j\in \{1,2,\ldots, dim(\hil)\}$)
which is ``right unitary":
\beq
\sum_\alpha T^\alpha_j T^{\alpha *}_{j'}= \delta^{j'}_j
\;,
\label{eq:right-uni}\eeq
and which satisfies

\beq
\sum_j T^\alpha_j \sqrt{\lambda_j} | \phi^j\rangle = 
\sqrt{w_\alpha} | \psi^\alpha\rangle
\;.
\label{eq:t-phi}\eeq
\end{theo}
proof:

($\Leftarrow$) Multiply each side of Eq.(\ref{eq:t-phi}) by its complex
conjugate and sum over $\alpha$.

($\Rightarrow$) Using Eqs.(\ref{eq:evec-expan}) and (\ref{eq:alpha-expan}) 
and the fact that the eigenvectors $| \phi^j\rangle$ are orthonormal, we get:

\beq
\sum_\alpha 
\sqrt{w_\alpha}
\langle\phi^j | \psi^\alpha \rangle
\langle\psi^\alpha | \phi^{j'} \rangle
\sqrt{w_\alpha} =
\lambda_j \delta^{j'}_j
\;.
\label{eq:hj-t-tstar}\eeq
For those $j$ such that $\lambda_j\neq 0$,
define $T$ by

\beq
T^\alpha_j = \sqrt{\frac{w_\alpha}{\lambda_j}} \langle \phi^j | \psi^\alpha\rangle
\;.
\label{eq:hj-t-def}\eeq
One can represent $T^\alpha_j$ as a matrix with rows labelled by $j\in \{1,2,\ldots, dim(\hil)\}$
 and columns labelled by $\alpha \in S_\rvalp$.
Eq.(\ref{eq:hj-t-def}) defines only those rows of $T$ with index $j$ such that $\lambda_j\neq 0$.
Eq.(\ref{eq:hj-t-tstar}) tells us that those rows 
which are defined by Eq.(\ref{eq:hj-t-def})
are orthonormal. The remaining rows of $T$
can be filled in using the  Gram-Schmidt
process \cite{Noble}. Once $T$ is fully specified, 
all the rows of $T$ are orthonormal, and therefore Eq.(\ref{eq:right-uni}) follows.
It is easy to check that the $T$ we have constructed also satisfies Eq.(\ref{eq:t-phi}). QED

The HJW Theorem refers to density matrices $\rho$ 
in an arbitrary Hilbert space $\hil$. 
But what if $\hil$ is a tensor 
product of two Hilbert spaces $\hilx$ and $\hily$?
Refs.\cite{Holland} and \cite{Heifei}
apply the HJW Theorem to tensor product spaces.
They prove the following theorem.

Consider a $\rho\in \dm(\hilxy)$ with eigenvalues  $\lambda_j$ and 
corresponding eigenvectors $|\phi^j\rangle$ for all $j\in N_{\rvx\rvy}$.
Let 

\beq
\Lambda = diag(\lambda_1, \lambda_2, \ldots, \lambda_{N_{\rvx\rvy}})
\;.
\eeq
For any matrix $M_{jj'}$ with $j,j' \in S_{\rvx \rvy}$, define

\beq
\langle M\rangle_{\alpha \beta} = \sum_{j, j'} T^\alpha_j M_{j, j'} T^{\beta *}_{j'}
\;,
\eeq
for all $\alpha, \beta\in S_\rvalp$.

\begin{theo}\label{th:quan-nec-suf-cond1}
$\rho\in\dm(\hilxy)$ is separable
if and only if there exists a matrix $T^\alpha_j$ ($\alpha\in S_\rvalp$, $j\in S_{\rvx\rvy}$)
and a set of vectors $\{ | \psi^\alpha \rangle \in \hilxy | \alpha\in S_\rvalp\}$
which satisfy:

\beq
| \psi^\alpha \rangle = 
| \psi^\alpha_\rvx \rangle 
| \psi^\alpha_\rvy \rangle 
\;,
\eeq
where $| \psi^\alpha_\rvx \rangle\in \hilx$ and $| \psi^\alpha_\rvy \rangle\in \hily$,

\beq
w_\alpha = \av{\Lambda}_{\alpha\alpha}
\;,
\eeq

\beq
\rho = \sum_\alpha w_\alpha \ket{\psi^\alpha } \bra{\psi^\alpha} = 
\sum_\alpha w_\alpha \pi(\psi_\rvx^\alpha)\pi(\psi_\rvy^\alpha)
\;,
\eeq

\beq
\sum_\alpha T^\alpha_j T^{\alpha *}_{j'}= \delta^{j'}_j
\;,
\eeq

\beq
\sum_j T^\alpha_j \sqrt{\lambda_j} | \phi^j\rangle = 
\sqrt{w_\alpha} | \psi^\alpha\rangle
\;.
\eeq

\end{theo}
proof:

We postpone proving this theorem since its proof follows from the proof of the following theorem. QED

The last theorem 
is a very powerful tool because it parametrizes 
with a linear transformation $T$
the space one must search to find a 
decomposition of a separable state $\rho$.
Besides the constraint that $T$ be right unitary, the theorem imposes no other constraints on the search space.
In particular, it avoids imposing inequality
constraints on the search space which other methods might impose in order
to enforce the positivity of the eigenvalues of $\rho$, $\pi(\psi_\rvx^\alpha)$,
$\pi(\psi_\rvy^\alpha)$.

Although the last theorem is very powerful, 
it is somewhat distant from classical considerations. 
One wonders whether
one can find a set of 
necessary and sufficient conditions for quantum separability 
that  more closely resemble the set of necessary and sufficient conditions for
classical separability that we gave in Theorem \ref{th:clas-separ}.
Indeed one can, as the following theorem shows.

Define the following array of operators:

\beq
[K_{\rvx\rvy}]_{j,j'} = \sqrt{\lambda_j} |\phi^j\rangle \langle\phi^{j'}| \sqrt{\lambda_{j'}}
\;,
\label{eq:kxy-def}\eeq
for $j,j'\in S_{\rvx\rvy}$.
The matrix elements of $K_{\rvx\rvy}$ with respect to the $|xy\rangle$ basis will be denoted by:

\beq
[K_{xy; x'y'}]_{j,j'} = [\langle xy|K_{\rvx\rvy}|x'y'\rangle]_{j,j'} = \sqrt{\lambda_j} \phi^j_{xy} \phi^{j'*}_{x'y'}\sqrt{\lambda_{j'}}
\;.
\label{eq:kxyxy-def}\eeq
We also define a partial trace of  $K_{\rvx\rvy}$ with respect to $\rvy$:

\beq
K_\rvx = \tr_\rvy K_{\rvx\rvy}
\;,
\eeq
whose matrix elements in the $|x\rangle$ basis are

\beq
K_{xx'}= \langle x|K_\rvx|x'\rangle
\;.
\eeq
Analogously, $K_{\rvy}$ and 
$K_{yy'}$ will stand for the partial trace of $K_{\rvx\rvy}$ with respect to $\rvx$,
and the matrix elements thereof.

\begin{theo}\label{th:quan-nec-suf-cond2}
$\rho\in \dm(\hilxy)$ is separable
if and only if there exists a transformation $T^\alpha_j$ 
($\alpha\in S_\rvalp$, $j\in S_{\rvx\rvy}$)
which is right unitary:
\beq
\sum_\alpha T^\alpha_j T^{\alpha *}_{j'}= \delta^{j'}_j
\;,
\label{eq:right-uni2}\eeq
and satisfies

\beq
\av{ K_{xy; x'y'}}_{\alpha\alpha} 
\av{ \Lambda}_{\alpha\alpha} 
=
\av{ K_{x, x'}}_{\alpha\alpha} 
\av{ K_{y, y'}}_{\alpha\alpha} 
\;
\label{eq:quan-cond-indep}\eeq
for all $x,x'\in S_\rvx$, $y,y'\in S_\rvy$ and $\alpha\in S_\rvalp$.
\end{theo}
proof:

($\Rightarrow$)
Since $\rho$ is separable, there exits $w_\alpha \in \pd(S_\rvalp)$,
and for each $\alpha\in S_\rvalp$, there exist states
$\ket{\psi^\alpha_\rvx}\in \hilx$,
$\ket{\psi^\alpha_\rvy}\in \hily$ so that

\beq
\rho = \sum_\alpha w_\alpha \pi(\psi^\alpha_\rvx) \pi(\psi^\alpha_\rvy) 
\;.
\eeq
$\rho$ can also be expanded in terms of its eigenvalues and eigenvectors:

\beq
\rho = \sum_j  \lambda_j \ket{\phi^j}\bra{\phi^j}
\;.
\eeq
Equating these two expressions for $\rho$ and taking matrix elements in 
the eigenvector basis gives:

\beq
\sum_\alpha 
\sqrt{w_\alpha}
\av{ \phi^j | \psi^\alpha_\rvx \psi^\alpha_\rvy}
\av{ \psi^\alpha_\rvx \psi^\alpha_\rvy  | \phi^{j'} }
\sqrt{w_\alpha}
=
\lambda_j \delta_j^{j'}
\;.
\label{eq:t-tstar}\eeq
For those $j$ such that $\lambda_j\neq 0$,
define $T$ by

\beq
T^\alpha_j = \sqrt{\frac{w_\alpha}{\lambda_j}} \langle \phi^j | \psi^\alpha_\rvx \psi^\alpha_\rvy\rangle
\;.
\label{eq:t-def}\eeq
One can represent $T^\alpha_j$ as a matrix with rows labelled by $j\in S_{\rvx\rvy}$ and columns labelled by $\alpha\in S_\rvalp$.
Eq.(\ref{eq:t-def}) defines only those rows of $T$ with index $j$ such that $\lambda_j\neq 0$.
Eq.(\ref{eq:t-tstar}) tells us that the rows 
defined by Eq.(\ref{eq:t-def}) 
are orthonormal. The remaining rows of $T$
can be filled in using the  Gram-Schmidt
process \cite{Noble}. Once $T$ is fully specified, all the rows of $T$ are orthonormal, 
so it is right unitary. Plugging the $T$ matrix just constructed into the definition 
Eq.(\ref{eq:kxyxy-def}) for $\av{K_{xy;x'y'}}_{\alpha \alpha}$ yields

\begin{subequations} \label{eq:kxyxy-and-traces}
\beq
\av{K_{xy;x'y'}}_{\alpha \alpha} = w_\alpha
\psi^\alpha_x \psi^\alpha_y
\psi^{\alpha *}_{x'} \psi^{\alpha *}_{y'}
\;.
\eeq
Thus

\beq
\av{K_{x,x'}}_{\alpha \alpha} = 
\sum_y \av{K_{xy;x'y}}_{\alpha \alpha} = w_\alpha \psi^\alpha_x \psi^{\alpha *}_{x'}
\;,
\eeq

\beq
\av{K_{y,y'}}_{\alpha \alpha} = 
\sum_x \av{K_{xy;xy'}}_{\alpha \alpha} = w_\alpha \psi^\alpha_y \psi^{\alpha *}_{y'}
\;,
\eeq

\beq
\av{\Lambda}_{\alpha \alpha} = 
\sum_{x,y} \av{K_{xy;xy}}_{\alpha \alpha} = w_\alpha
\;.
\eeq
\end{subequations}
Eqs.(\ref{eq:kxyxy-and-traces}) clearly imply 
Eq.(\ref{eq:quan-cond-indep}).

($\Leftarrow$) 
Summing Eq.(\ref{eq:kxyxy-def}) for $\av{K_{xy;x'y'}}_{\alpha \alpha}$ over $\alpha$
and using the right unitarity of $T$ yields

\beq
\sum_\alpha 
\av{K_{xy;x'y'}}_{\alpha \alpha}  =  \sum_j \lambda_j \phi^j_{xy} \phi^{j*}_{x'y'} = \av{xy |\rho| x'y'}
\;.
\eeq
We define $w_\alpha$ for all $\alpha$ by

\beq
w_\alpha = \av{\Lambda}_{\alpha \alpha}
\;.
\eeq
Using the last two equations, we get

\beq
\av{xy |\rho| x'y'}
=
\sum_\alpha
\av{K_{xy;x'y'}}_{\alpha \alpha}
=
\sum_\alpha
w_\alpha
\av{x | \rho^\alpha_\rvx | x'}
\av{y | \rho^\alpha_\rvy | y'}
\;,
\eeq
where, for all $\alpha$ such that $w_\alpha\neq 0$, 
$\rho^\alpha_\rvx$ and $\rho^\alpha_\rvy$ are defined by

\beq
\av{x | \rho^\alpha_\rvx | x'} = \frac{\av{K_{xx'}}_{\alpha \alpha}}{w_\alpha}
\;,\;\;
\av{y | \rho^\alpha_\rvy | y'} = \frac{\av{K_{yy'}}_{\alpha \alpha}}{w_\alpha}
\;.
\eeq
Clearly, $\rho^\alpha_\rvx\in \dm(\hilx)$ and $\rho^\alpha_\rvy\in \dm(\hily)$.
QED

In the section on classical separability, we plotted $\tilde{P}(x,y,\alpha)$ at
each point $(x,y,\alpha)\in S_{\rvx\rvy\rvalp}$ of a 3-dimensional orthogonal lattice with 
axes $x,y,\alpha$. We noted that for a separable $P(x,y)$, its $\tilde{P}(x,y,\alpha)$ 
is a product-corrugated surface on each $\alpha$ plane. Theorems \ref{th:quan-nec-suf-cond1}
and \ref{th:quan-nec-suf-cond2} on quantum separability show that similar plots are possible
in the quantum case. One can plot the phase and magnitude of $\psi^\alpha_{xy}$. For a separable
$\rho$,  $\psi^\alpha_{xy} = \psi^\alpha_{x}\psi^\alpha_{y}$, 
so both the phase and magnitude of $\psi^\alpha_{xy}$ are corrugated surfaces on each $\alpha$ plane. The magnitude is product
corrugated and the phase is mod-$2\pi$ addition corrugated. 
Note that $\sum_{x,y} |\psi^\alpha_{xy}|^2 = 1$, so $|\psi^\alpha_{xy}|^2$
summed over all points of any $\alpha$ plane gives one. Note also that
$A^\alpha_{xy} = \sqrt{w_\alpha} \psi^\alpha_{xy}$ satisfies 
$\sum_{x,y, \alpha} |A^\alpha_{xy}|^2 = 1$, so 
$|A^\alpha_{xy}|^2$
summed over all lattice points is one.

\begin{theo}\label{th:bridge}
Suppose $\rho\in \dm(\hilxy)$
can be  expanded thus:

\beq
\rho = \sum_\alpha w_\alpha \rho^\alpha
\;,
\eeq
where $w_\alpha \in \pd(S_\rvalp)$, and $\rho^\alpha\in \dm(\hilxy)$ for all $\alpha\in S_\rvalp$.
Furthermore, suppose $\{ \ket{\alpha} | \alpha\in S_\rvalp\}$ is an orthonormal basis 
of $\hilalp$ and that $\sigma\in \hilxyalp$ is defined by

\beq
\sigma = \sum_\alpha w_\alpha \ket{\alpha}\bra{\alpha} \rho^\alpha
\;.
\label{eq:sigma-def}\eeq
(Note that $\rho = \tr_\alpha \sigma$). Then

\beq
S_\sigma(\rvx : \rvy | \rvalp) = \sum_\alpha w_\alpha S_{\rho^\alpha}(\rvx : \rvy)
\;.
\label{eq:bridge}\eeq

\end{theo}
proof:

By definition,

\beq
S_\sigma(\rvx : \rvy | \rvalp) =
S_\sigma (\rvx, \rvalp)
+S_\sigma (\rvy, \rvalp)
-S_\sigma (\rvx, \rvy, \rvalp)
-S_\sigma (\rvalp)
\;.
\label{eq:cmi-def}\eeq
Each of the terms on the right hand side can be broken into two parts. Consider
for example the  $S_\sigma (\rvx, \rvalp)$ term:

\begin{subequations}\label{eq:s_decomps}
\begin{eqnarray}
\lefteqn{S_\sigma (\rvx, \rvalp) =}\nonumber\\
&=& -\tr_{\rvx,\rvalp}[ \tr_\rvy(\sigma) \log \tr_\rvy(\sigma)] =\nonumber\\
&=& -\sum_\alpha \tr_{\rvx}[ \tr_\rvy(w_\alpha \rho^\alpha) \log \tr_\rvy(w_\alpha \rho^\alpha)]=\nonumber\\
&=& H(\vec{w}) + \sum_\alpha w_\alpha S_{\rho^\alpha}(\rvx)
\;, 
\end{eqnarray}
where $H(\vec{w})$ is the classical entropy for the probability distribution $\{w_\alpha | \alpha\in S_\rvalp\}$.
Likewise, one can show that

\beq
S_\sigma (\rvy, \rvalp) =  H(\vec{w}) + \sum_\alpha w_\alpha S_{\rho^\alpha}(\rvy)
\;,
\eeq

\beq
S_\sigma (\rvx, \rvy, \rvalp) =  H(\vec{w}) + \sum_\alpha w_\alpha S_{\rho^\alpha}(\rvx,\rvy)
\;,
\eeq

\beq
S_\sigma (\rvalp) =  H(\vec{w})
\;.
\eeq
\end{subequations}
Plugging Eqs.(\ref{eq:s_decomps}) into the right hand side 
of Eq.(\ref{eq:cmi-def}) establishes Eq.(\ref{eq:bridge}). QED

See Ref.\cite{Tucci-tang99} to learn how to build quantum Bayesian nets which yield a
density matrix like the $\sigma$ (see Eq.(\ref{eq:sigma-def})) in the last theorem.

Suppose $\rho \in \dm(\hil)$ can be expressed as 
$\rho = \sum_\alpha w_\alpha \ket{\psi^\alpha}\bra{\psi^\alpha}$,
where $w_\alpha \in \pd(S_\rvalp)$ and  $\ket{\psi^\alpha}\in \hil$ for all $\alpha$.
Then we say the set  ${\cal E} = \{ (w_\alpha, \ket{\psi^\alpha}) | \alpha\in S_\rvalp\}$
is a { \it $\rho$ ensemble}. In particular, 
the set of pairs of eigenvalues and corresponding eigenvectors of $\rho$
constitutes a $\rho$ ensemble which we will denote by ${\cal E}_0$ and call the
{\it standard $\rho$ ensemble}. Eq.(\ref{eq:t-phi}) of the HJW Theorem  
can be represented  schematically by 
$T {\cal E}_0 = {\cal E}$.

For any $\rho\in\dm(\hilxy)$, the {\it entanglement of formation} is defined by

\beq
E_F(\rho) = \min_{\cal E} \sum_\alpha w_\alpha S[ \tr_\rvy (\ket{\psi^\alpha} \bra{\psi^\alpha})]
\;,
\label{eq:ef-stan-def}\eeq
where the minimum is taken over the set of all $\rho$ ensembles 
${\cal E} = \{ (w_\alpha, \ket{\psi^\alpha}) | \alpha\in S_\rvalp\}$. 
But the HJW Theorem
taught us that any $\rho$ ensemble  ${\cal E}$
can be parametrized by 
a right unitary matrix $T$ such that $T {\cal E}_0 = {\cal E}$.
 Thus, we can also define
$E_F(\rho)$ as a minimum over all right unitary matrices $T^\alpha_j$ 
with $\alpha\in S_\rvalp$ and $j\in S_{\rvx\rvy}$.
Furthermore, 
if we define $\rho^\alpha = \pi(\psi^\alpha)$ for all $\alpha$, then 
$S[ \tr_\rvy \pi(\psi^\alpha)] = S_{\rho^\alpha}(\rvx)$.
Thus, Eq.(\ref{eq:ef-stan-def})
can be rewritten as 

\beq
E_F(\rho) = \min_{T} \sum_\alpha w_\alpha  S_{\rho^\alpha}(\rvx)
\;.
\label{eq:quant-inter-eform}\eeq
But observe that $\rho_\alpha$ is a pure state of $\dm(\hilxy)$ so that
$S_{\rho^\alpha}(\rvx) = S_{\rho^\alpha}(\rvy)$ and $S_{\rho^\alpha}(\rvx, \rvy) = 0$
so $S_{\rho^\alpha}(\rvx) = \frac{1}{2}S_{\rho^\alpha}(\rvx:\rvy)$.  
Using this observation, Eq.(\ref{eq:quant-inter-eform}) and Theorem \ref{th:bridge}, we get 

\beq
E_F(\rho) 
= \frac{1}{2}\min_{T} \sum_\alpha w_\alpha  S_{\rho^\alpha}(\rvx:\rvy)
= \frac{1}{2}\min_{T} S_{\sigma}(\rvx: \rvy | \rvalp)
\;,
\label{eq:ef-and-cmi}\eeq
where $\sigma = \sum_\alpha w_\alpha \ket{\alpha}\bra{\alpha} \rho^\alpha$.

\begin{theo}
$\rho\in \dm(\hilxy)$ is separable
if and only if 
$E_F(\rho) =0$.
\end{theo}
proof:

($\Rightarrow$) If $\rho$ is separable then 
$\rho = \sum_\alpha w_\alpha \rho^\alpha$, where 
$\rho^\alpha = \pi(\psi_\rvx^\alpha) \pi(\psi_\rvy^\alpha)$. Thus,
$\sum_\alpha w_\alpha  S_{\rho^\alpha}(\rvx:\rvy) = 0$. 

($\Leftarrow$) If $E_F(\rho) =0$ then 
there exist a right unitary matrix $T$ and a $\rho$ ensemble ${\cal E}$ such that
$T {\cal E}_0 = {\cal E}$. If 
${\cal E} = \{ (w_\alpha, \ket{\psi^\alpha})| \alpha\in S_\rvalp\}$, then
$\rho = \sum_\alpha w_\alpha \rho^\alpha$, $\rho^\alpha = \pi(\psi^\alpha)$ and
$ S_{\rho^\alpha}(\rvx:\rvy) =0$ for all $\alpha$. Because its mutual entropy vanishes,
$\rho^\alpha = \rho^\alpha_\rvx \rho^\alpha_\rvy$
where $\rho^\alpha_\rvx\in \dm(\hilx)$ and 
$\rho^\alpha_\rvy\in \dm(\hily)$. Thus, $\rho$ is separable. QED

\section{Similarities}

The previous two sections have discussed classical(C) and quantum(Q) separability. 
We end the paper by discussing the following table, which 
enumerates some of the similarities between the two cases:

\begin{center}
{\renewcommand{\arraystretch}{1.3}
{\footnotesize
\begin{tabular}{|l|l|l|}\hline
 			& Classical & Quantum \\ 
\hline\hline
The unknown: & $\tilde{P}(x,y, \alpha)$ & $T^\alpha_j$ \\
\hline
Boundary conditions  		 & $\sum_\alpha \tilde{P}(x, y, \alpha) = P(x,y)$ & $\sum_\alpha T^\alpha_j T^{\alpha*}_{j'} = \delta^{j'}_j$ \\
satisfied by the unknown:    &                                                &          											  \\
\hline
Integral equations	& $\tilde{P}(x,y,\alpha) \sum_{x_1,y_1}\tilde{P}(x_1,y_1,\alpha) =$ & 
$\av{K_{xy;x' y'}}_{\alpha\alpha}\sum_{x_1,y_1}\av{K_{x_1 y_1;x_1 y_1}}_{\alpha\alpha}=$ \\
satisfied by the unknown:   & $\sum_{y_1}\tilde{P}(x,y_1,\alpha)\sum_{x_1}\tilde{P}(x_1,y,\alpha)$ & 
$ \sum_{y_1}\av{K_{x y_1;x' y_1}}_{\alpha\alpha} \sum_{x_1}\av{K_{x_1 y;x_1 y'}}_{\alpha\alpha} $\\                                            										
\hline
Entropic eqn. equivalent	& $H(\rvx : \rvy | \rvalp) = 0 $ & $S_\sigma(\rvx:\rvy|\rvalp)=0$ \\
to boundary value prob.:    &    for prob. dist. $\tilde{P}(x,y,\alpha)$ &  for $\sigma = \sum_\alpha w_\alpha \ket{\alpha}\bra{\alpha} \rho^\alpha$\\                                             											  \hline
\end{tabular}
}
}
\end{center}

We proved a theorem that says that  C separability of $P(x,y)\in \pd(S_{\rvx \rvy})$ implies the existence
of a certain  ``unknown" $\tilde{P}(x,y, \alpha)\in \pd(S_{\rvx \rvy\alpha})$.
Likewise, we proved a theorem that says that Q separability of $\rho\in \dm(\hilxy)$ implies the existence
of a certain  ``unknown"  $T^\alpha_j$.
In both the C and Q cases, the unknown must satisfy certain constraints which can
be thought of as representing a boundary value problem comprising a set of
discrete integral equations with boundary conditions.
In both the C and Q cases, the existence of a solution to the 
boundary value problem was proven to be equivalent to 
the statement that a certain conditional mutual information vanishes.

\end{document}